\documentclass[aps,prl,preprint,superscriptaddress]{revtex4-1} %prb draft
\usepackage{graphicx}
\usepackage{xcolor}
\usepackage{hyperref}
\hypersetup{
    colorlinks=false,
    citebordercolor=[rgb]{0,0,1}
}
\usepackage{array}
\newcolumntype{C}[1]{>{\centering\let\newline\\\arraybackslash\hspace{0pt}}m{#1}}
\begin{document}

%\preprint{}

\title{Epitaxial stabilization of thin films of the frustrated Ge-based spinels}


\author{Denis M. Vasiukov}
\email[]{vasyukov@physics.msu.ru}
\affiliation{Department of Physics and Astronomy, Rutgers University, Piscataway, New Jersey 08854, USA}
\author{Mikhail Kareev}
\affiliation{Department of Physics and Astronomy, Rutgers University, Piscataway, New Jersey 08854, USA}
\author{Fangdi Wen}
\affiliation{Department of Physics and Astronomy, Rutgers University, Piscataway, New Jersey 08854, USA}
\author{Liang Wu}
\affiliation{Department of Physics and Astronomy, Rutgers University, Piscataway, New Jersey 08854, USA}
\author{Padraic Shafer}
\affiliation{Advanced Light Source, Lawrence Berkeley National Laboratory, Berkeley, CA 94720, USA}
\author{Elke Arenholz}
\affiliation{Advanced Light Source, Lawrence Berkeley National Laboratory, Berkeley, CA 94720, USA}
\affiliation{Cornell High Energy Synchrotron Source, Cornell University, Ithaca, NY 14853, USA}
\author{Xiaoran~Liu}
\affiliation{Department of Physics and Astronomy, Rutgers University, Piscataway, New Jersey 08854, USA}
\author{Jak Chakhalian}
\affiliation{Department of Physics and Astronomy, Rutgers University, Piscataway, New Jersey 08854, USA}
%\thanks{}


\date{\today}

\begin{abstract}
Frustrated magnets can host numerous exotic many-body quantum and topological phenomena. GeNi$_2$O$_4$ is a three dimensional $S=1$ frustrated magnet with an unusual two-stage transition to the two-dimensional antiferromagnetic ground state, while GeCu$_2$O$_4$ is a high-pressure phase with a strongly tetragonally elongated spinel structure and magnetic lattice formed by  $S=1/2$~CuO$_2$~linear chains with frustrated interchain exchange interactions and exotic magnetic behavior. Here we report on the first thin-film epitaxial stabilization of these two compounds.  Developed growth mode, surface morphology, crystal structure and copper valence state were characterized by \emph{in-situ} reflection high-energy electron diffraction, atomic force microscopy, X-ray reflectivity, X-ray diffraction, X-ray photoelectron spectroscopy and resonant X-ray absorption spectroscopy. Our results pave an alternative route to the comprehensive investigation of the puzzling magnetic properties of these compounds and exploration of novel emergent features driven by strain.
\end{abstract}

%\pacs{75.30.Wx, 76.80.+y, 61.50.Ks}
%\keywords{}

\maketitle


Research on magnetic systems possessing frustration, low dimensionality or their combination, is a very active and fruitful subfield with a great potential for emergent phenomena, new states of matter and exotic excitations exemplified by spin ice, quantum spin liquids, and spin-charge separation to name a few~\cite{lacroix2011introduction,castelnovo2012spin,savary2016quantum,zhou2017quantum}. Despite a plethora of interesting theoretical proposals the mapping of theory to realistic material systems still remains a formidable challenge, particularly, due to demand for high-quality materials that can host such exotic phenomena and states~\cite{savary2016quantum,broholm2020quantum}. As one of the alternatives to the solid-state chemistry routes, ultra-thin films may be a viable option to address the challenge.

%In particular, one-dimensional (1D) spin chains can be mapped into the simplest theoretical models yet with several surprising phenomena, e.g. the Haldane conjectures\cite{haldane19883}, i.e the fundamental difference between integer (spin-gapped) and half-integer (spin-gapless) spin chains\cite{zhou2017quantum}, and unusual ballistic spinnon driven thermal transport along the chains\cite{sologubenko2007thermal}.

Among the magnetically frustrated materials, complex oxides with the spinel structure are of special interest. Spinel oxides have the general formula AB$_2$O$_4$, composed of a cubic close-packed sublattice of anions in which 1/8 of the tetrahedral (A-site) and 1/2 of the octahedral (B-site) interstices are filled by cations. In the absence of distortions, spinel crystallizes in the cubic space group $Fd\bar{3}m$ and the B-site cations form a network of corner-shared tetrahedra (also termed as ``pyrochlore'') sublattice that can potentially trigger the strongest frustration in three dimensions~\cite{takagi2011highly}. Following this direction, the so-called ``4--2'' spinel family with germanium on the A-site is particularly interesting (general formula Ge$^{4+}$B$^{2+}_{2}$O$_{4}$), as this class of compounds has magnetically inactive A-site; hence, their magnetic properties are solely determined by the pyrochlore sublattice.

In this work, we report on the first time epitaxially stabilized thin films of two members of the ``4--2'' spinel family, GeNi$_2$O$_4$~(GNO) and GeCu$_2$O$_4$~(GCO), which exhibit very different low-dimensional magnetic behavior. The bulk GNO undergoes a peculiar two-stage phase transition in to a 2D magnetic ground state~\cite{crawford2003structure,stevens2004heat,diaz2006magnetic,lancaster2006magnetism,matsuda2008frustrated,lashley2008specific}, while the bulk GCO shows a multiferroic transition with interacting 1D $S=1/2$ chains\cite{yamada2000spin,zou2016up,zhao2018multiferroicity,yanda2018spontaneous}.

\begin{figure*}[t]
\includegraphics[width=16.5cm, keepaspectratio=true]{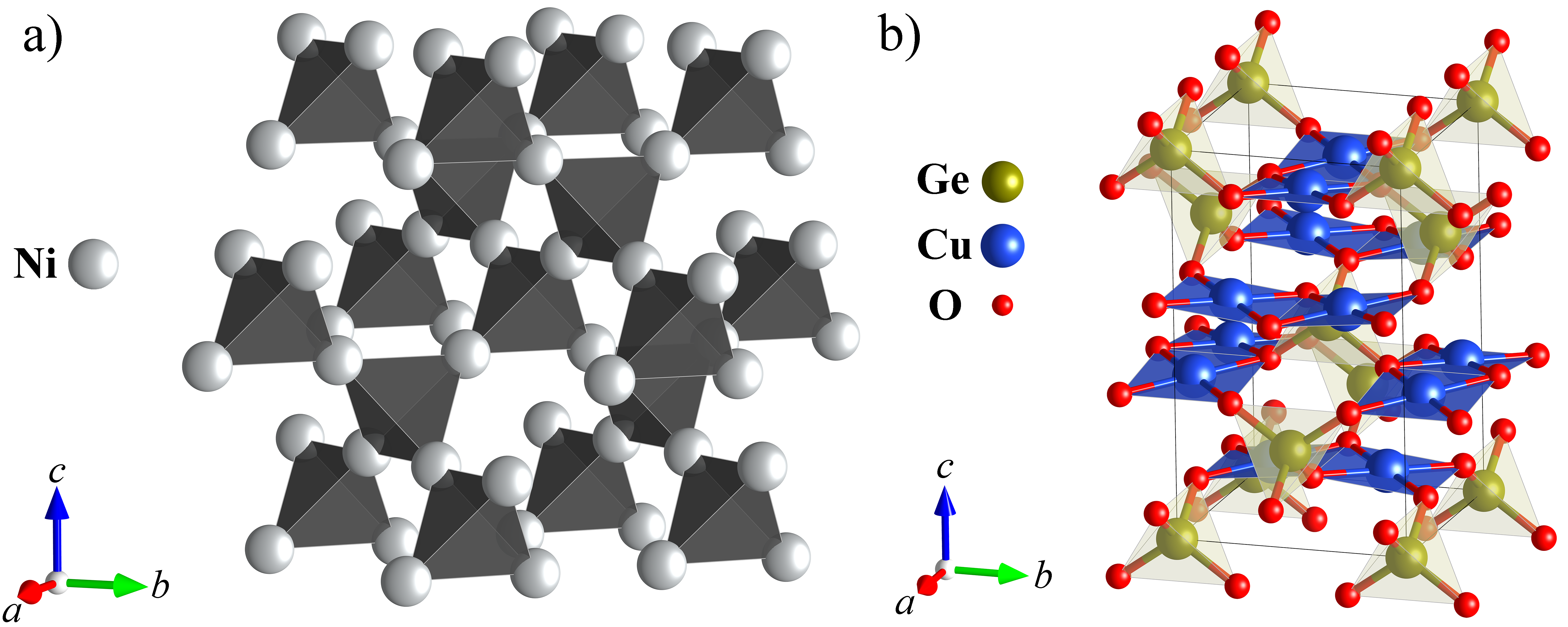}
\caption{\label{Figure_1} Crystal structures of GeNi$_{2}$O$_{4}$ and  GeCu$_{2}$O$_{4}$ spinels. a) In GNO the B-site sublattice of Ni$^{2+}$ ions form an ideal pyrochlore sublattice of corner-shared tetrahedra, the Ge and O ions are omitted for clarity. b) The large tetragonal elongation in GCO results in the effective square-planar coordination of Cu$^{2+}$ ions therefore the B-site sublattice can be regarded as a set of alternating mutually perpendicular layers of 1D chains. These structure figures were prepared using VESTA~3 software~\cite{momma2011vesta}.}
\end{figure*}

In GNO the B-site Ni$^{2+}$ has the $3d^8$ electronic configuration, which adopts a non-degenerate electronic term $^3A_{2g}$ with negligible single-ion anisotropy. The non-degeneracy of Ni$^{2+}$ indicates that frustration cannot be relieved via structural distortion driven by the cooperative Jahn-Teller effect making this pyrochlore lattice of isotropic $S=1$ spins, shown in Fig.~\ref{Figure_1}a, a promising candidate for unusual magnetic phenomena. Indeed, study of NaCaNi$_2$F$_7$ revealed continuum of quantum fluctuations characteristic for Coulomb phase with only spin-glass-like freezing transition~\cite{plumb2019continuum}. In contrast to this case, the unusual sequence of magnetic transitions is observed in bulk GNO at low temperatures leading to antiferromagnetically ordered ground state with a propagation vector $q=(\frac{1}{2}\:\frac{1}{2}\:\frac{1}{2})$ corresponding to the rhombohedral lattice system~\cite{bertaut1964structure}. Upon this transition the B-site splits magnetically into alternating kagome and triangular planes along the (111)~direction, which form a set of intrinsically 2D frustrated sublattices.

More detailed recent studies show that this antiferromagnetic state is developed through two consecutive first-order phase transitions separated by $\sim0.7$~K ($T_{N1}=12.1$~K and $T_{N2}=11.4$~K) without any measurable structural distortion~\cite{crawford2003structure,stevens2004heat,diaz2006magnetic,barton2014structural}. A subsequent muon-spin relaxation experiment suggested that these two transitions correspond to distinct ordering of two magnetic subsystems~\cite{lancaster2006magnetism}. Based on the single-crystal neutron elastic scattering data~\cite{matsuda2008frustrated} it was concluded that the first transition ($T_{N1}$) corresponds to a magnetic ordering only within the kagome planes (each plane has ferromagnetic ordering stacked antiferromagnetically) while the triangular planes become magnetically ordered at the second transition ($T_{N2}$). Strikingly, the saturated magnetic moment of the triangular planes remains two times smaller than one of the kagome planes~\cite{diaz2006magnetic,matsuda2008frustrated,basu2020magnetic}, although in the absence of distortions all Ni$^{2+}$ ions are crystallographically equivalent. A strong magnetic anisotropy of the ground state was demonstrated in the single-crystal experiments with applied magnetic field~\cite{hara2004field,basu2020magnetic}. This highly unusual magnetic behavior seems to be closely related to the {G}d$_2${T}i$_2${O}$_7$ with pyrochlore structure which shows similar magnetic transitions~\cite{stewart2004phase}.

The specific-heat data indicate several puzzling features, which remain unexplained. These include the presence of substantial magnetic correlations in the paramagnetic state despite the low frustration factor ($f=\Theta_{CW}/T_N\sim0.7$), the coexistence of gapped and gapless spin waves, and missing of $\sim40$~\% of the expected magnetic entropy~\cite{lashley2008specific}. Furthermore, unexpectedly the change in magnetic entropy is almost equal for both phase transitions~\cite{stevens2004heat,lashley2008specific}, which clearly contradicts to the picture of two separate orderings in kagome and triangular planes (as triangular lattice contain three times less Ni$^{2+}$ ions relative to the kagome lattice).

As for the second synthesized spinel, GCO is a recoverable high-pressure phase with the hausmannite structure type (space group $I4_1/amd$). Interestingly, it is only the second tertiary oxide discovered within the ternary Ge-Cu-O system~\cite{hegenbart1981high} while the first one is the celebrated GeCuO$_3$ with 1D antiferromagnetic $S=1/2$ chains, and the first discovered spin-Peirls transition (i.e., spin pairing with a valence bond formation) among the inorganic compounds~\cite{hase1993observation}. GCO has a tetragonally distorted spinel structure in which the octahedral B-site is occupied by the Cu$^{2+}$ Jahn-Teller active ions. Due to the large Jahn-Teller distortion, the local coordination of Cu$^{2+}$ ions can be described as almost square-planar with the in-plane Cu-O distance of 1.939~{\AA} and the apical oxygens located at 2.504~{\AA}~\cite{hegenbart1981high}. As shown in Fig.~\ref{Figure_1}b, such a large distortion allows to consider this spinel structure as comprised of alternating mutually perpendicular layers of 1D CuO$_2$ $S=1/2$ chains interconnected via GeO$_4$ tetrahedra. The 1D nature of magnetic interactions within the $S=1/2$~chains is supported by magnetic susceptibility data that show the characteristic Bonner-Fisher behavior with the ratio of inter-chain to intra-chain exchange couplings $J^\prime/J\sim0.16$~\cite{yamada2000spin}. The susceptibility data also indicate the onset of long-range magnetic ordering at 33~K~\cite{yamada2000spin}.

Despite such a large tetragonal distortion of the pyrochlore sublattice, the magnetic interaction between Cu ions in GeCu$_2$O$_4$ remain frustrated in close analogy to the Cs$_2$CuCl$_4$ case, where the frustrated interchain coupling results in the exotic ``triplon'' bound state that readily moves between chains~\cite{kohno2007spinons}. Theoretically, such an anisotropic pyrochlore lattice was mapped onto a 2D crossed-chain model~\cite{starykh2005anisotropic}, which predicts the formation of a valence-bond solid with a crossed-dimer state in the limit of small $J^\prime/J$~\cite{starykh2005anisotropic}. However, contrary to the model calculations, the density-functional theory (DFT) study of GCO suggests the formation of a spiral magnetic ground state~\cite{tsirlin2011spiral}. The subsequent neutron powder diffraction experiment revealed the presence of collinear antiferromagnetism at low temperatures with an unexpected up-up-down-down (i.e. $\uparrow\uparrow\downarrow\downarrow$) spin ordering pattern along the chains~\cite{zou2016up}. This unusual ordering pattern was attributed to the presence of bi-quadratic exchange interaction active in this compound~\cite{zou2016up} while \emph{a posteriori} DFT study argued that actual reason is the negligibly small nearest-neighbor coupling~\cite{badrtdinov2019origin}. Surprisingly, two independent groups recently reported spin-induced multiferroicity in GCO emerging at $T_N\sim33$~K~\cite{zhao2018multiferroicity,yanda2018spontaneous}, which is incompatible with any theoretically proposed or experimentally determined magnetic structures~\cite{starykh2005anisotropic,tsirlin2011spiral,zou2016up}. To date, the nature of multiferroicity in GCO remains largely unknown.

Taking into account puzzling magnetism of these compounds and the fact that epitaxial strain can alter the underlying microscopic Hamiltonian of a system, fabricating thin films of GNO and GCO is of great interest since it can potentially shed light on the open questions about the magnetic behavior of the bulk crystals and may lead to new emergent quantum states. Moreover, in the bulk form GCO is only stable at high-pressure ($\sim$~4~GPa~\cite{hegenbart1981high}) and its high-pressure solid-state synthesis is difficult and yields only \emph{micron-sized} single crystals hindering application of many important probes. The epitaxial strain can help to stabilize phases which are otherwise unstable at ambient pressure in the bulk~\cite{gorbenko2002epitaxial}, therefore exploring the fabrication of GCO thin films can offer alternative opportunities for realizing large-area single crystals.

Here, we report on the first successful fabrication of GNO and GCO thin films by means of pulsed laser deposition. The developed growth mode, surface morphology, film thickness  and crystal structure were characterized by \emph{in-situ} reflection high-energy electron diffraction (RHEED), atomic force microscopy (AFM), X-ray reflectivity (XRR) and X-ray diffraction (XRD). Stoichiometry of the films and ion valency were investigated by X-ray photoelectron spectroscopy (XPS). Electronic states in the GNO film were further studied by synchrotron based resonant X-ray absorption spectroscopy (XAS) on Ni L-edge and O K-edge. The combination of advanced tools confirms the successful epitaxial growth of high-quality (001)-oriented GCO and GNO thin films.
%The XPS spectra were measured using a Thermo Scientific K-Alpha XPS spectrometer with monochromated Al $K_{\alpha}$ radiation. The beam spot size was 400~$\mu$m$^2$ and we used flood gun for charge compensation.

\paragraph*{\textbf{Stabilization of the GeNi$_2$O$_4$ thin film.}}

We begin with the description of details of GNO growth. Taking into account that the bulk magnetic ground state in GNO has $q=(\frac{1}{2}\:\frac{1}{2}\:\frac{1}{2})$, it is of particular interest to obtain an epitaxially strained film grown in the direction  which is noncollinear to this propagation vector. For this reason we have chosen (001)-oriented MgAl$_2$O$_4$ (MAO) substrate with spinel structure for the deposition.

The UV-laser (KrF excimer laser with 248~nm wavelength) was operated at 3~Hz repetition rate with 2.7~J/cm$^2$ energy density per pulse to ablate a stoichiometric GeNi$_2$O$_4$ target. All films were deposited on the substrate of $5\times5$~mm$^2$ area with average surface roughness of $S_a\sim100$~pm heated to 700~$^{\circ}$C as measured by a pyrometer under 6 mTorr of pure oxygen. After deposition, samples were annealed at the growth condition and then cooled down to room temperature with the initial rate of 15 $^\circ$C/min under the same oxygen pressure. All experimental data presented below were collected using the same film.

\begin{figure}[t]
\includegraphics[width=7cm, keepaspectratio=true]{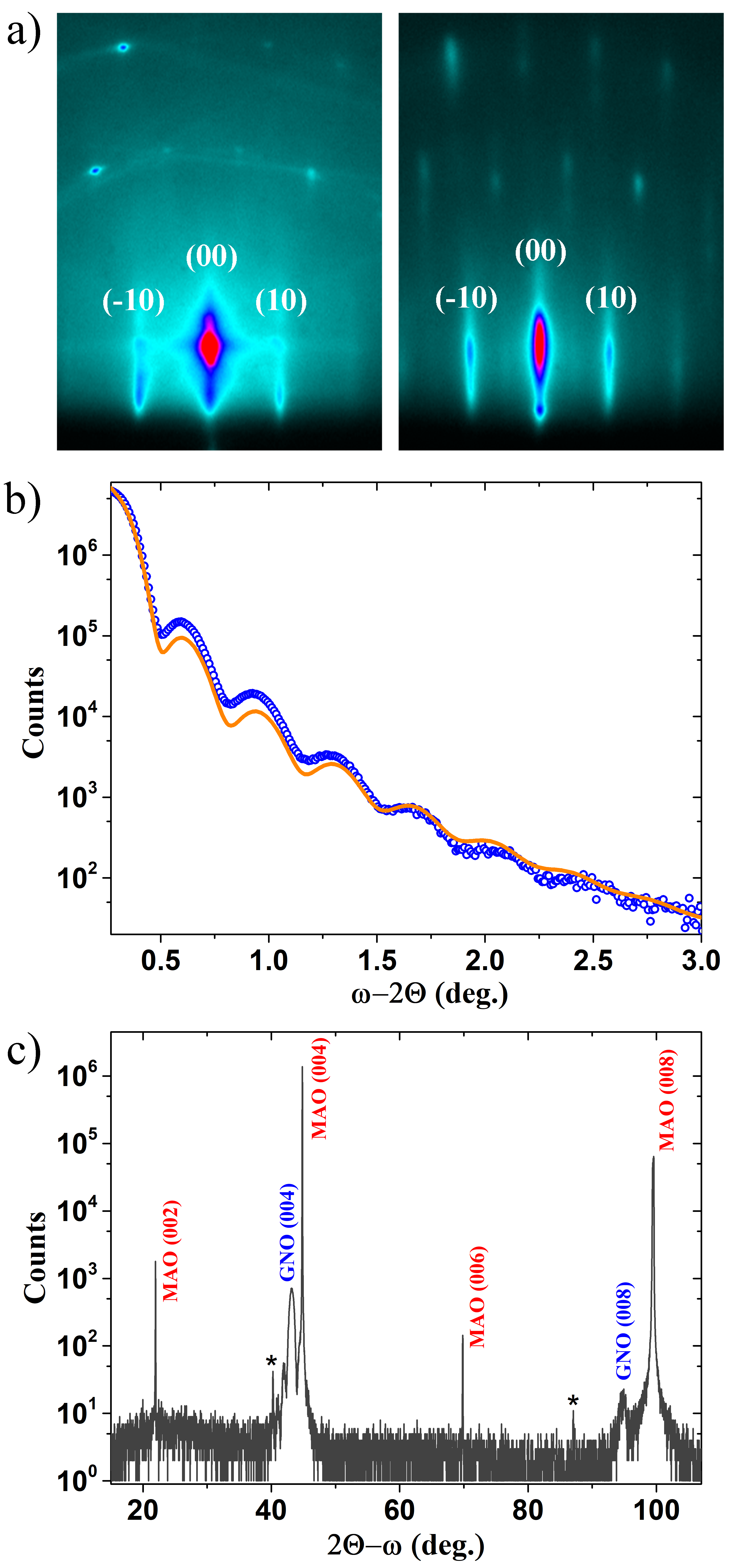}
\caption{\label{Figure_2} Characterization of the grown GNO thin film. a)~On the left the RHEED pattern of (001)-oriented MAO substrate at 650~$^\circ$C and on the right the RHEED pattern of the grown GNO film at room temperature. b)~The fit of XRR curve for the same sample yields 11.79(4)~nm film thickness and 240(20)~pm surface roughness. c)~$2\Theta$-$\omega$ X-ray diffraction scan along $00l$ direction. The scan contains allowed reflections of the substrate and fabricated film together with two forbidden reflections (002 and 006) of the substrate which arise because of the Umweganregung effect~\cite{renninger1937umweganregung}. Asterisks mark (004) and (008) peaks of the substrate due to residual $K_{\beta}$ radiation.}
\end{figure}

\begin{figure*}[t]
\includegraphics[width=16.5cm, keepaspectratio=true]{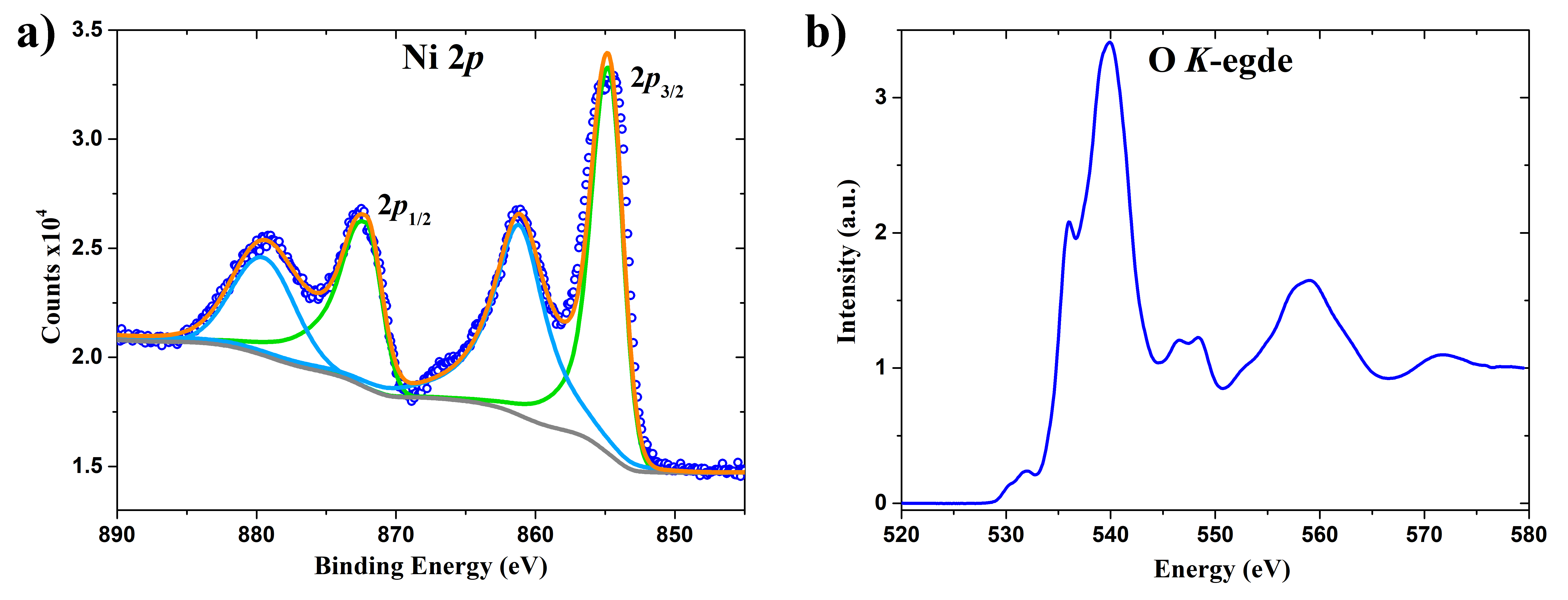}
\caption{\label{Figure_3} XPS and oxygen $K$-edge XAS spectra of GNO thin films. a) High resolution XPS scan of Ni $2p$ states shows the spin-orbit split doublet (green) together with satellite features (light blue). The background line is shown in gray color. Unsplitted doublet confirms that film contain Ni only in a single valence state. b) Oxygen $K$-edge XAS spectrum demonstrates absence of the sharp pre-peak feature that signifies the $d^8$ electron configuration of Ni$^{2+}$ ions without significant admixture of configurations with oxygen $p$-holes (\emph{e.g.} $d^9L$).}
\end{figure*}

The deposition process was monitored \emph{in-situ} by high-pressure RHEED. The representative RHEED patterns with indexed zero-order Laue zone of the substrate and fabricated film are presented in Fig.~\ref{Figure_2}a. As expected for two isostructural compounds the patterns are similar including identical of higher order Laue zones. The intensity of the specular reflection shows only an initial reduction at the start of the deposition, followed by a rapid recovery that remains constant without any evident oscillations during further deposition. The well-developed streak pattern shown in Fig.~\ref{Figure_2}a indicates that the GNO film grow in the step-flow mode~\cite{hasegawa2002reflection}.

X-ray reflectivity and diffraction measurements were performed with Empyrean diffractometer using the Cu~$K_\alpha$~radiation. Fit of XRR data (Fig.~\ref{Figure_2}b) yields a surface roughness $S_a=240(20)$~pm, which is in a good agreement with the AFM results ($\sim200$~pm); both XRR and AFM data confirm the development of smooth surface morphology of the GNO film. In addition, fitting results of the XRR data yield a film thickness of 11.79(4)~nm. The $2\Theta-\omega$ scan shown in Fig.~\ref{Figure_2}c contains reflections of both the substrate (red) and two reflections of the GNO film (blue). These peaks are indexed as (004) and (008) reflections, confirming the (001)-orientation of the film with no other secondary chemical phases present. As the bulk lattice parameter of GNO is $a=8.22$~{\AA}~\cite{hirota1990study}, growth on the MAO substrate ($a=8.08$~{\AA}) should result in 1.7~\% in-plane compressive strain, elongating the out-of-plane lattice constant. Indeed, from our diffraction data we have determined that the out-of-plane lattice constant is 8.376(5)~{\AA}, which is consistent with compressive strain.

The XPS spectra were measured by a Thermo Scientific K-Alpha XPS spectrometer with monochromated Al~$K_\alpha$~radiation. Stoichiometry of the GNO film was determined in high-resolution core-shell scans around $2p$ state of Ge and Ni and $1s$ state of O. The analysis of XPS data results in Ge~:~Ni~:~O~=~1~:~1.98~:~4.24 which is consistent within the experimental uncertainty with the desired GeNi$_2$O$_4$ composition (the deviation in the oxygen value from the ideal stoichiometric ratio is due to the surface exposed to ambient). Figure~\ref{Figure_3}a shows a characteristic core-shell Ni~$2p$ scan. As seen, the energy position of the Ni $2p_{3/2}$ peak at 856.5~eV and the absence of any peak splitting imply expected single valency of Ni in the film.

\begin{figure}[b]
\includegraphics[width=8.3cm, keepaspectratio=true]{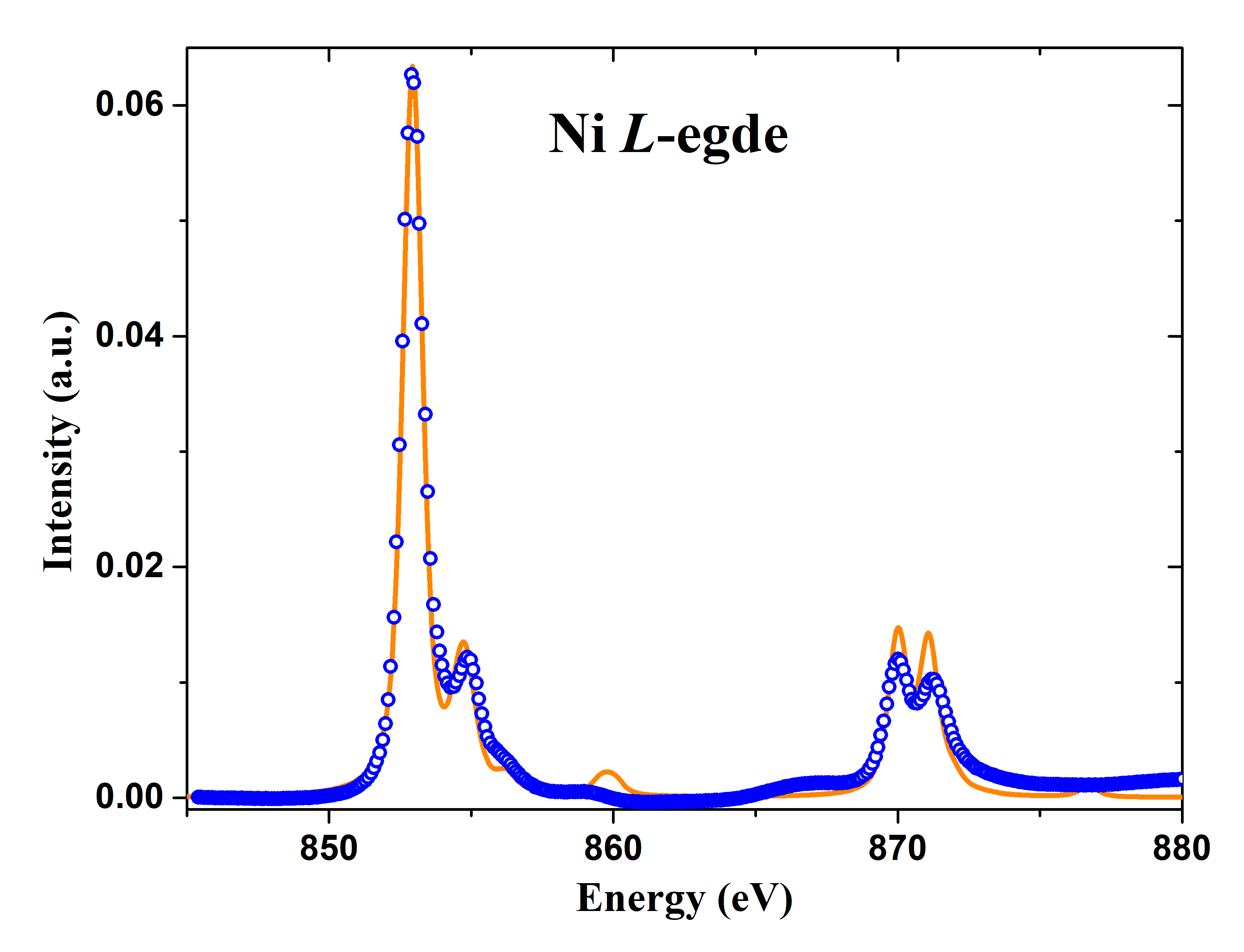}
\caption{Ni $L$-edge XAS spectrum of the GNO film. Blue dots are the measured data whereas orange line show simulated spectrum of Ni$^{2+}$.\label{Figure_4}}
\end{figure}

To further investigate the electronic state of Ni~and O~away from the surface, we carried out XAS experiments at beamline~4.0.2 of the Advanced Light Source (Lawrence Berkeley National Laboratory). Oxygen $K$-edge scan shown in Fig.~\ref{Figure_3}b is very similar to the theoretical line-shape~\cite{osti_1200493} and does not show any sizable pre-edge intensity around 529~eV, reflecting only a small contribution of configurations with ligand holes (e.g.~$3d^9L$) to the electronic state of Ni~ion. Furthermore, in accord with the XPS data the Ni $L$-edge line-shape corresponds to Ni in $2+$ state. The simulated spectrum of Ni$^{2+}$ shown in Fig.~\ref{Figure_4} corroborates the divalent nature of Ni~ions\footnote{Simulation was done using Crispy software version 0.7.2 developed by Marius Retegan. In the simulation we used 0.9~reduction factor of default Slater integrals for Ni$^{2+}$ in the software and the parameters $U_{3d,3d}$ and $U_{2p,3d}$ were 7.3 and 8.5~eV, respectively.}. An optimization of the calculated line-shape to the experimental data yields the following atomic parameters: crystal field splitting $10Dq=0.9$~eV, charge-transfer gap of 4~eV, hopping integrals $V_{e_g}=2.5$~eV and $V_{t_{2g}}=1.0$~eV; those values are in agreement with the values of Ni$^{2+}$ reported for NiO~\cite{haverkort2012multiplet}.

\begin{figure}[t]
\includegraphics[width=7.5cm, keepaspectratio=true]{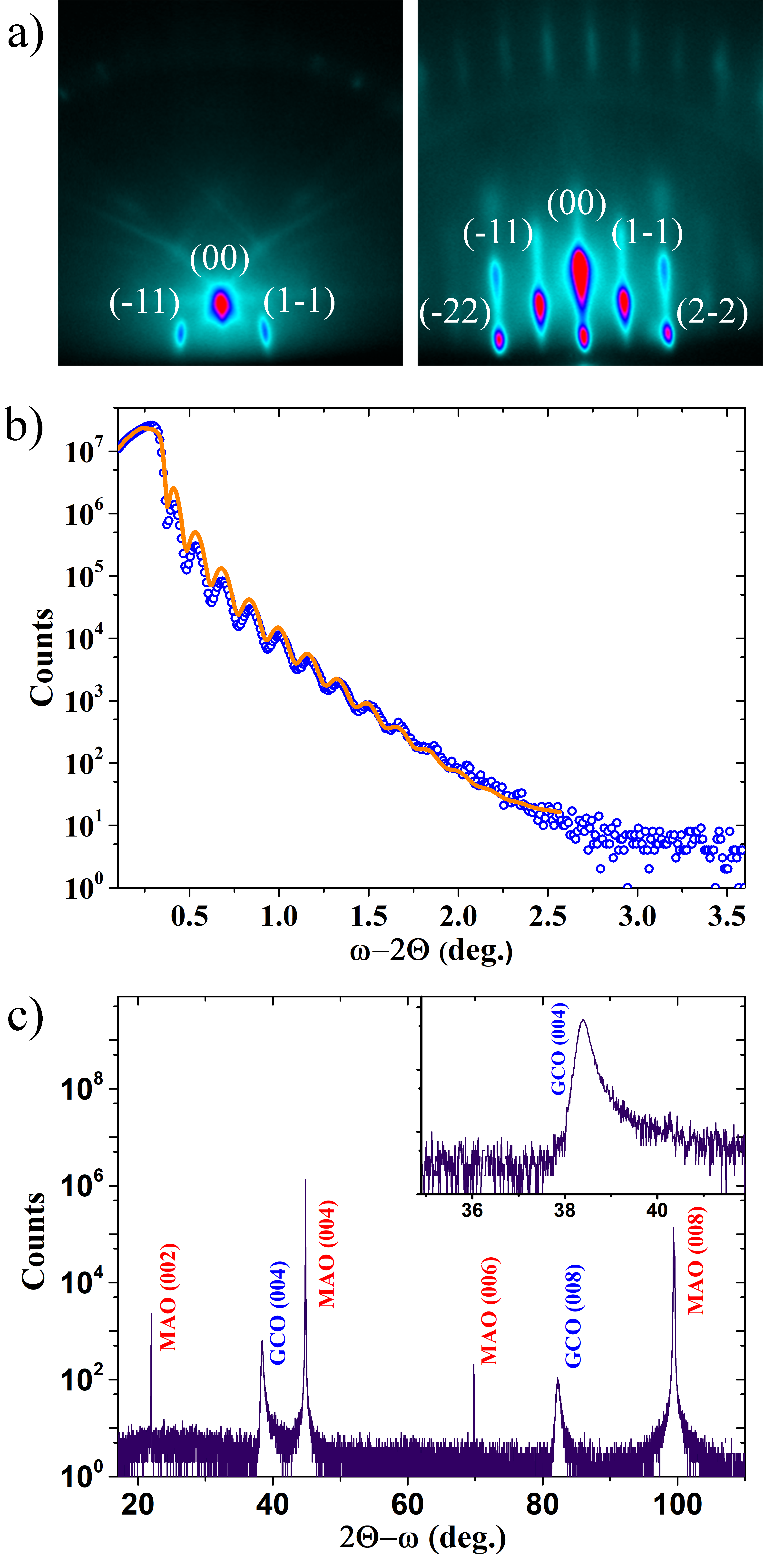}
\caption{\label{Figure_5} Characterization of the grown GCO thin films. a)~The RHEED patterns of MAO substrate and GCO film at 600~$^\circ$C are on the left and right, respectively. b)~The fit of XRR curve of this film yields 26~nm film thickness. c)~$2\Theta$-$\omega$ X-ray diffraction scan along $00l$ direction of another GCO film with 50~nm thickness. The scan contains allowed reflections of the substrate and fabricated film together with two forbidden reflections (002 and 006) of the substrate which arise because of the Umweganregung effect\cite{renninger1937umweganregung}. The inset shows asymmetric shape of (004) GCO peak which indicates relaxation of the substrate-induced strain.}
\end{figure}

\paragraph*{\textbf{Stabilization of the GeCu$_2$O$_4$ thin film.}}

In the case of GCO, the choice of a suitable substrate is not obvious. Since the bulk GeCu$_2$O$_4$ is stable only above 4~GPa~\cite{hegenbart1981high}, it is natural select a substrate offering compressive strain. However, due to the very large tetragonal elongation along $c$-axis, GCO lattice is strongly compressed in the $ab$ plane, leading to no commercially available crystals with the same spinel structure to provide suitable compressive strain in the $ab$ plane~\cite{hill1979systematics}. As a compromise, we selected two substrates for our experiments: (\emph{i})~(001)-oriented SrTiO$_3$ (STO) with cubic perovskite structure, which would result in compressive strain, and (\emph{ii})~(001)-oriented MAO spinel substrate, which is isostructural to GCO but provides moderate \emph{tensile} strain.

For the extensive series of depositions on (001)-oriented STO we varied both substrate temperatures from 470 to 700~$^\circ$C and background oxygen pressure from 5 to 100 mTorr. The GCO phase failed to stabilize under any of these growth conditions, and only a Cu$_2$O phase with Cu ions in +1~oxidation state readily forms at substrate temperatures above 600~$^\circ$C and low oxygen pressures.

\begin{figure*}[t]
\includegraphics[width=16.5cm, keepaspectratio=true]{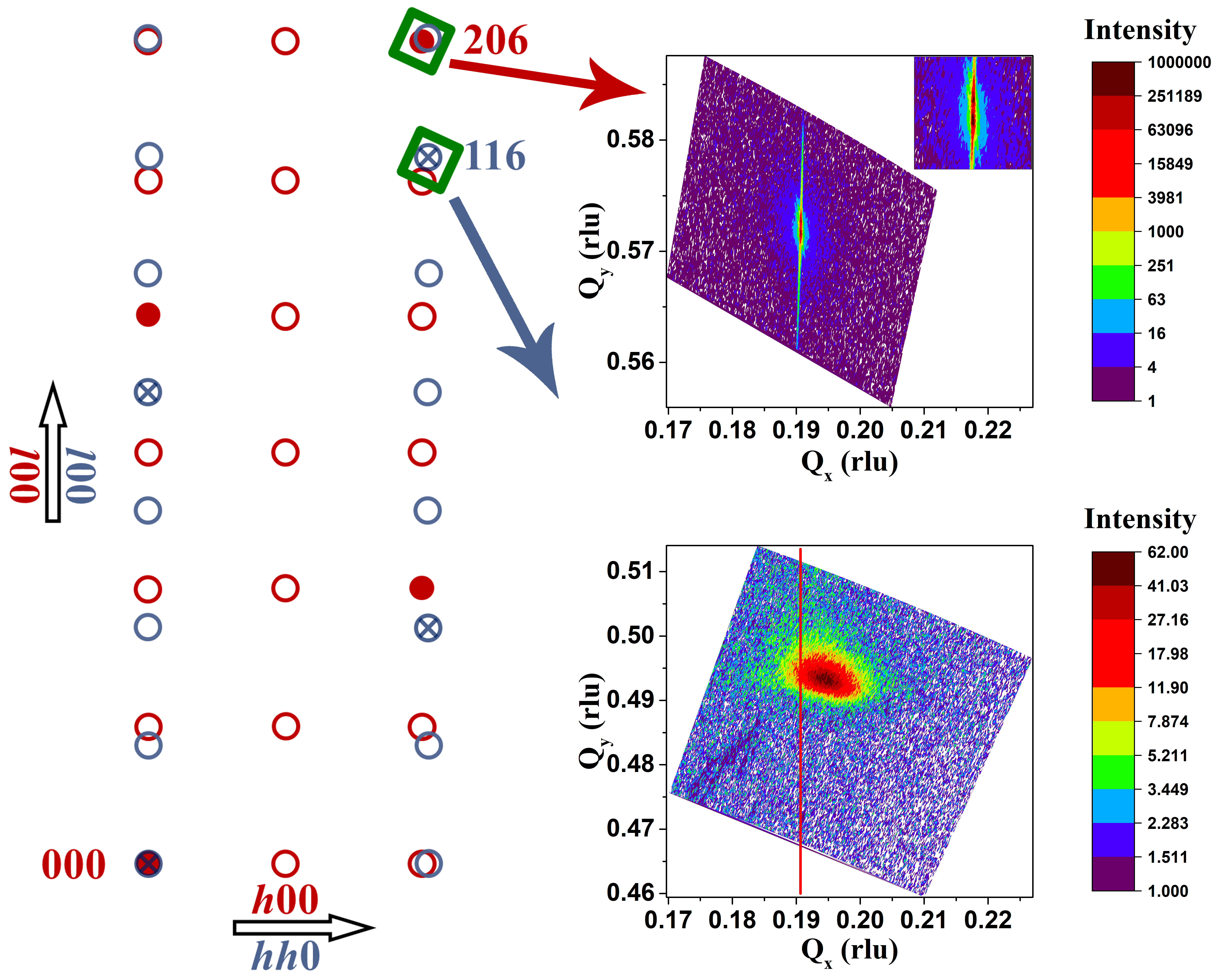}
\caption{Reciprocal space mapping of the 50~nm~thick GCO film. On the left the reciprocal lattices of MAO substrate together with GCO film are drawn for the case of (001) epitaxial growth. The filled and open red circles correspond to allowed and forbidden reflections of the MAO substrate, respectively, while blue circles designate reflections of the GCO film; in this case allowed reflections are marked by crossed circle. The orientation of the scans coincide with drawn reciprocal lattice i.e. $Q_x$ and $Q_y$ are parallel to $h00$ and $00l$ directions of the substrate reciprocal lattice, respectively. The top-right figure is a scan around the 206 substrate reflection. The vertical stripe is due to non-monochromaticity of laboratory Cu $K_\alpha$ source. The inset shows zoomed 206 peak where one can see both $K_{\alpha 1}$ and $K_{\alpha 2}$ doublet peaks. The bottom-right figure is a scan around the 116 film reflection. The vertical red line designates the $Q_x$ coordinate of the substrate peak.\label{Figure_6}}
\end{figure*}

Next, we repeated a series of depositions on the (001)-oriented MAO substrate. At low oxygen pressure of $\sim5$~mTorr and in the temperature range $600-700$~$^\circ$C, all the samples stabilize in to Cu$_2$O phase, akin to the growth on STO substrate at these conditions. However, upon increasing the oxygen pressure above 50~mTorr, the GeCu$_2$O$_4$ phase appears to be stabilized within the temperature range of $600-700$~$^\circ$C. Further optimization of the growth condition confirmed that the best quality GCO films can be obtained at 50 mTorr of O$_2$ and 600~$^{\circ}$C substrate temperature. After deposition all samples were annealed at the growth condition and then gradually cooled down (15 $^\circ$C/min) to room temperature at the same oxygen pressure. The presented experimental data below were collected using two GCO films with 26~and~50~nm thicknesses.

The entire growth process was monitored by \emph{in-situ} high-pressure RHEED. The representative RHEED pattern of the 26~nm thick GCO film is shown in Fig.~\ref{Figure_5}a. The intensity of the specular reflection shows only an initial reduction at the start of the deposition followed by intensity recovery. As the growth sequence progressed, it remains nearly constant without any evident oscillations. The elongated RHEED streaks with intensity modulation in Fig.~\ref{Figure_5}a indicate the presence of a multilevel stepped surface~\cite{hasegawa2002reflection}. This leads to the relatively high average surface roughness of 630(10)~pm as determined from the XRR data presented in Fig.~\ref{Figure_5}b. The AFM scans confirmed the presence of the stepped surface with roughness of separate terraces as low as $S_a\sim100-200$~pm, which is similar to the quality of the MAO substrate surface.

\begin{figure*}[t]
\includegraphics[width=16.5cm, keepaspectratio=true]{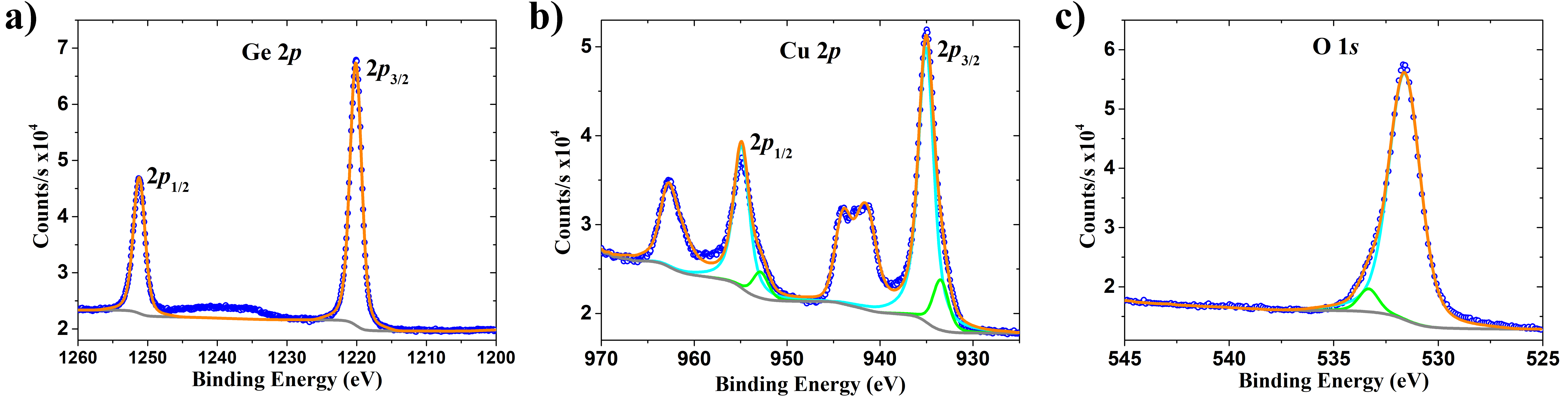}
\caption{\label{Figure_7} High resolution core-shell XPS scans of the 26~nm~thick GCO thin film: a) Ge $2p$ shell, b) Cu $2p$ shell and c) O $1s$ shell. The copper scan shows mostly Cu$^{2+}$ (light-blue lines) with developed satellite features. The minor amount of Cu$^{1+}$ and different state of O (green lines) is likely related to the surface decomposition of the film.}
\end{figure*}

Our data also indicate that the 2D growth mode of GCO gradually turns into the 3D island mode as the thickness of film increases. This is evident by the occurrence of spotty (3D transmission like) patterns on the RHEED images for a 50~nm thick film\footnote{In this case the film thickness was determined from the number of laser pulses.}. This growth mode leads to the films with rough surface that is consistent with absence of oscillations in the XRR spectrum of this film.

Next we discuss the results of our diffraction experiments.  The $00l$ scan of the 50~nm thick GCO film shown in Fig.~\ref{Figure_5}c contains two reflections of the GCO film (blue) apart from the substrate reflections (red). These peaks can be indexed as the (004) and (008) reflections of GCO, confirming the (001)-orientation and absence of secondary phases. To compare the out-of-plane lattice parameter of the GCO film with the bulk, we convert the lattice parameters of MAO from the face-centered unit cell to the equivalent body-centered one ($a_I=a_F/\sqrt2$). Since the bulk GCO lattice parameters are $a=5.593$~{\AA} and $c=9.395$~{\AA}~\cite{hegenbart1981high}, for the MAO substrate ($a_F=8.08$~{\AA}) it would result in 2.15~\% in-plane tensile strain, shortening the out-of-plane lattice constant. Indeed, our diffraction data for the 25~nm and 50~nm thick films yield the out-of-plane value of 9.31~{\AA} and 9.37~{\AA}, respectively, consistent with the expected value of tensile strain. Note, the difference in the lattice constant for 25 and 50~nm thick films implies a different degree of strain relaxation, which is also evident from the asymmetric shape of (004) GCO peak in the inset of Fig.~\ref{Figure_5}c.

To further validate that the film is indeed the targeted GeCu$_2$O$_4$ phase, we carried out reciprocal space mappings (RSM) on the same 50~nm film. A schematic illustration for the reflections in the reciprocal space is plotted in Fig.~\ref{Figure_6}, on which the allowed and forbidden reflections of both GCO and MAO are shown. Here we note that since the out-of-plane lattice parameter of the film significantly differs from the substrate (9.37~{\AA}~\emph{vs}~8.08~{\AA}), the peak positions of the substrate and film diverge considerably in the reciprocal space. As the result, one can barely perform a usual RSM which includes both substrate and film reflections together. Instead, we measured two separate scans around the MAO (206) peak and the expected position of GCO (116) peak.

Indeed, as seen in Fig.~\ref{Figure_6} the film peak was found in the predicted position, which further demonstrates the stabilization of the desired GCO phase. Critically, the measured $Q_x$ of GCO (116) reflection is 0.1945(5)~rlu, corresponding to the in-plane lattice parameter of~5.62~{\AA}. This value is larger than 5.59~{\AA} of bulk GCO, but smaller than 5.71~{\AA} of MAO substrate. Therefore, it implies an expected partial relaxation of strain inside the 50~nm film.

We further verify the Cu oxidation state and the stoichiometry of film by carrying out XPS experiment on the 26~nm GCO film (see Fig.~\ref{Figure_7}). The scan around Cu $2p$ core state shows that a majority of copper ions have the expected 2+ oxidation state (935~eV $2p_{3/2}$ peak) with the well-developed satellite features shown in Fig.~\ref{Figure_7}b. However, a small feature seen at 933.5~eV ($\sim9$~\% of total Cu signal) implies the presence of Cu$^{1+}$ state. This observation is consistent with the fact that our high-resolution scans around $2p$ state of Ge and Cu yield Ge~:~Cu~$\approx$~1~:~1.2 which is also in variance with the bulk-like GeCu$_2$O$_4$ composition.
Since our XRD data do not show the presence of any additional peaks, in agreement with the absence of granulation in AFM scans common for multi-phase films, we attribute the small admixture of Cu$^{1+}$ and deviation in the cation ratio to the instability of the surface termination layer and/or degradation of the air-exposed surface of GCO films.

In summary, for the first time we have successfully fabricated single-crystalline epitaxial thin-films of (001)-oriented GeNi$_2$O$_4$ with the $S=1$ pyrochlore sublattice and the high-pressure phase of GeCu$_2$O$_4$ with the network of $S=1/2$ linear chains. A combination of advanced spectroscopic and diffraction techniques confirmed high structural and chemical quality of the films. The synthesis of coherently strained (001) GeNi$_2$O$_4$ thin films can provide additional opportunities for the study of unusual magnetic transition in the $S=1$ pyrochlore lattice and their response to the epitaxial strain. In addition, the availability of large area GeCu$_2$O$_4$ thin films opens a road towards detailed experimentation to reveal the controversial nature of its ground state magnetism and elucidate the origin of multiferroicity in this compound.

\begin{acknowledgments}
The work in Rutgers University was supported by the Gordon and Betty Moore Foundation's EPiQS Initiative through Grant No.~GBMF4534, and by the Department of Energy under Grant No.~DE-SC0012375. This research used resources of the Advanced Light Source, which is a Department of Energy Office of Science User Facility under Contract No.~DE-AC0205CH11231.
\end{acknowledgments}

\bibliography{Bibliography/My_bib}

\end{document}